\documentclass[fleqn,usenatbib]{mnras}
\usepackage{newtxtext,newtxmath}
\usepackage[T1]{fontenc}
\usepackage{ae,aecompl}
\usepackage{graphicx} % Including figure files\
\usepackage{amsmath} % Advanced maths commands\
\usepackage{color}
\newcommand{\kms}{km\,s$^{-1}$}

\newcommand{\beii}{Be\,{\sc ii}}
\newcommand{\bevii}{$^{7}$Be}
\newcommand{\beviiii}{$^{7}$Be\,{\sc ii}}

\newcommand{\lii}{Li\,{\sc i}}

\newcommand{\livii}{$^{7}$Li}
\newcommand{\liviii}{$^{7}$Li\,{\sc i}}

\newcommand{\cai}{Ca\,{\sc i}}
\newcommand{\caii}{Ca\,{\sc ii}}
\newcommand{\caiii}{Ca\,{\sc iii}}
\newcommand{\nai}{Na\,{\sc i}}
\newcommand{\ki}{K\,{\sc i}}
\newcommand{\mgi}{Mg\,{\sc i}}
\newcommand{\mgii}{Mg\,{\sc ii}}
\newcommand{\hi}{H\,{\sc i}}
\newcommand{\hei}{He\,{\sc i}}

\newcommand{\iiihe}{$^{3}$He}

\newcommand{\nii}{N\,{\sc ii}}

\newcommand{\oii}{O\,{\sc ii}}
\newcommand{\oiii}{O\,{\sc iii}}

\newcommand{\feii}{Fe\,{\sc ii}}
\newcommand{\fei}{Fe\,{\sc i}}

\newcommand{\crii}{Cr\,{\sc ii}}

\newcommand{\neiii}{Ne\,{\sc iii}}
\newcommand{\nev}{Ne\,{\sc v}}
\begin{document}
% Title of the paper, and the short title which is used in the headers.\
% Keep the title short and informative.\
\title[\bevii\ detection in  nova]{  \bevii\ in the outburst  of the ONe nova   V6595 Sgr}
\author[]{Molaro, P. $^{1,2}$\thanks{E-mail: paolo.molaro@inaf.it }\thanks{Based on  data from Paranal Observatory, ESO, Chile}, Izzo, L.$^{3}$, D'Odorico, V $^{1}$, Aydi, E $^{4}$, Bonifacio, P.$^{5}$, Cescutti, G. $^{1,2,9}$
\newauthor
Harvey, E.J. $^{6}$, Hernanz, M. $^7$, Selvelli, P.$^{1}$, della Valle, M. $^8$ \\  
 $^{1}$  INAF-Osservatorio Astronomico di Trieste, Via G.B. Tiepolo 11, I-34143 Trieste, Italy\\
 $^{2}$  Institute of Fundamental Physics of the Universe, Via Beirut 2, Miramare,   Trieste, Italy\\
  $^{3}$ DARK, Niels Bohr Institute, University of Copenhagen, Jagtvej 128, 2200 Copenhagen, Denmark\\
  $^{4}$Center for Data Intensive and Time Domain Astronomy, Department of Physics and Astronomy, Michigan State University, East Lansing, MI 48824, USA\\
 $^{5}$ GEPI, Observatoire de Paris, Universit{\'e} PSL, CNRS, Place Jules Janssen, 92195 Meudon, France\\
  $^{6}$ Astrophysics Research Institute, Liverpool John Moores University, Liverpool, L3 5RF, UK \\
 $^7$ Institute of Space Sciences (ICE, CSIC) and IEEC, Campus UAB, Cam{\'i} de Can Magrans s/n, 08193 Cerdanyola del Valles (Barcelona), Spain\\
 $^8$ Capodimonte Astronomical Observatory, INAF-Napoli, Salita Moiariello 16, 80131-Napoli, Italy\\
 $^9$ INFN, Sezione di Trieste, Via A. Valerio 2, I-34127 Trieste, Italy}

\date{Accepted.... Received 2018...}
\pagerange{\pageref{firstpage}--\pageref{lastpage}} \pubyear{2002}
\maketitle
\label{firstpage}
\begin{abstract}
We  report  the search for \beviiii~  isotope in the outbursts of the classical nova V6595 Sgr by means of high resolution   UVES  observations taken at the ESO VLT in April 2021,  about two weeks after  discovery and under difficult circumstances due to the pandemic.  Narrow absorption  components with velocities at $\sim$  -2620 and -2820 \kms,  superposed on broader and shallow absorption,  are observed in  the  outburst spectra for the  \beviiii\   $\lambda\lambda$ 313.0583, 313.1228 nm  doublet  resonance lines,  as well as in  several other elements  such as \caii, \fei, \mgi, \nai,  {H\,{\sc i}} but \lii.   Using \caii~ K line as a reference element, we infer    
$N(\mbox{\bevii})/N(\mbox{H})$  
$\approx$ 7.4 $\cdot 10 ^{-6}$, or  $\approx$ 9.8 $\cdot 10 ^{-6}$  when the \bevii\ decay is taken into account. The \bevii\ abundance  is about half of the value most frequently measured in novae. 
The possible presence of over-ionization in the layers where \beviiii\ is detected is also discussed.
 Observations taken at the Telescopio Nazionale Galileo (TNG) in La Palma  91 days after discovery showed  prominent emission lines of Oxygen and Neon which allow  to  classify the nova  as  ONe type. Therefore, although \bevii\ is expected to be   higher in CO novae, it is found     at comparable levels in both nova types.

\end{abstract}
\section{Introduction}
Classical Novae (CNe) are recurring thermonuclear explosions  in binary-star systems formed by  a white dwarf (WD) accreting matter from a main sequence or evolved companion \citep{BodeEvansbook,DellaValleIzzo2020}. The layer of accreted material grows in mass until the pressure at its bottom becomes sufficiently high ($> 2 \times 10^{19}$ dyne/cm$^2$) for the beginning of nuclear ignition of the p-p chain. Since the bottom of the accreted layer is electron degenerate, nuclear burning increases  the temperature  without  expansion that would inhibit further nuclear reactions. When the temperature becomes sufficiently high ($T > 10^7$ K), the energy source shifts to the CNO cycle and both temperature and energy release increase at a much faster rate. When  pressure and temperature at the bottom of the accreted layer exceed the degeneracy values, thermonuclear reactions (TNR) ignite  removing degeneracy and causing ejection of matter into the interstellar medium  \citep{Gallagher1978}. The subsequent expansion of the hot envelope at  velocities of the order of 500-3000 \kms\  is responsible for the initial brightness of the nova, which  can reach an absolute magnitude of $V \sim -9.5$ mag \citep{selvelli2019A&A...622A.186S}. 

Synthesis of 
lithium in novae explosions was first predicted in the 1970s \citep{Arnould1975,Starrfield1978}. Lithium is created in the thermonuclear runaway via the 
reaction $^3$He($\alpha$,$\gamma$)$^7$Be with $^7$Be decaying to $^7$Li via electron capture (half-life 53.22 days).
The production proceeds through the so-called Cameron-Fowler mechanism \citep{Cameron1955,CameronFowler1971} where \bevii\ has to be transported to cooler zones than where it is formed, with a time-scale shorter that its electron capture time, and  therefore preserved from destruction.

The suggestion  which dates back  to  the 70's  was  thwarted by the non detection of \livii\ in a bunch of novae \citep{Friedjung1979}.  After decades of observational failures to detect the \lii 670.8 nm line, the parent nucleus $^7$Be   was  detected in classical  novae \citep{Tajitsu2015,Tajitsu2016,Molaro2016,Izzo2018,molaro2020MNRAS.492.4975M,Arai2021ApJ...916...44A},  thanks to  high-resolution spectrographs capable to reach the atmospheric cut-off.  Following these detections, a search in high dispersion spectra of historical novae in the archival database of the space-based International Ultraviolet Explorer (IUE)   led to the identification of  the $^7$Be II resonance line  in nova  V838 Her, which had been overlooked  \citep{Selvelli2018}. The possible presence of $^7$Li 670.8 nm resonance line in nova V1369 Cen  \citep{Izzo2015} was also reported.
The detection of the short-lived  $^7$Be in the  spectra of novae shortly after outburst implies that this isotope is freshly created in the thermonuclear runaway (TNR) processes of the nova event.  \bevii\ decays with  capture of an internal K-electron and ends into an  ionized lithium, whose ground-state transitions are not observable in the optical range, thus explaining the general non-detection of neutral $^7$Li   \citep{Molaro2016}. 
%So far,   $^7$Be II has been detected in all novae, 
%with the exception of ASASSN-17hx (V612 Sct), which on the other hand was  proposed as  a peculiar object \citep{mason2020A&A...635A.115M}. 
The decay  of \bevii\  into an excited state of \livii\ produces a  high-energy line at 478 keV emitted during the de-excitation to the ground state of the fresh \livii\  produced in the \bevii\  electron-capture \citep{Clayton1981ApJ...244L..97C,Gomez1998}. Several unsuccessful attempts to detect the line with Gamma ray satellites have been performed \citep{Harris2001}. The  detection of the radioactive \bevii\ nuclei  in the nova outburst reopened the possibility of detecting   the 478 Kev line with INTEGRAL for  nearby novae. The   distance should be lower than $\approx$ 0.5 kpc, though the horizon would depend on the amount of \bevii\ produced in the nova event \citep{jean2000MNRAS.319..350J,Siegert2018,siegert2021arXiv210400363S}. 
%The nova ASASSN-18fv was observed with INTEGRAL-Director Discretionary Time, with a total exposure of 2.8 Ms. Although the complete data analysis is still ongoing, it is already clear that only upper limits to the 478 Kev emission line have been obtained \citep{siegert2021arXiv210400363S}.
We note that the detection of the 478 keV line from a nova with 
known distance will allow us to derive an accurate \bevii~ abundance.

The astrophysical origin  of lithium  represents a major open question in modern 
astrophysics \citep{Fields2011}.  The abundance of lithium  observed in halo  and metal-rich disk stars  is much lower and higher, respectively, than 
the  primordial value estimated from Big Bang nucleosynthesis once the baryonic density  from CMB or deuterium abundance is adopted. This implies   the existence of efficient lithium sinks and factories which are both still unknown. For the latter, many astrophysical 
sources can actively produce lithium such as  AGB stars \citep{Romano2001},  red giants \citep{Wallerstein1982},  classical novae \citep{Starrfield1978} or spallation processes \citep{Davids1970}.  Early Galactic chemical evolution models  of $^7$Li  revealed that classical novae show the required time scales, but lithium nova yields were theoretically  derived and not supported by observations at that time \citep{DAntonaMatteucci1991,Romano1999,Romano2001}.  The recent measured yields  imply   massive \bevii\ ejecta with the final \livii\  product  at about four  orders of magnitude  above meteoritic   abundance and novae are very likely the major Li sources in the Galaxy \citep{Molaro2016,Cescutti2019,Romano2021arXiv210611614R}. 
%With such an overproduction, novae could be  the most  important   $^7$Li factories \citep{Romano2021}.  

A major problem  is that the measured abundances exceed what foreseen by  current nova nucleosynthesis models  \citep{Hernanz1996,Jose1998}, and  it seems that \bevii\ is present in both fast and slow novae, with comparable abundances.  
  Controversy arises 
about the maximum amount of \beviiii\  that can be produced in novae (see e.g. \citealp{Deenissenkov2021MNRAS.501L..33D}, and discussions in \citealp{starrfield2020ApJ...895...70S}) but it remains well below what derived from observations.  \citet{chugai2020} suggested
that the presence   of an over-ionization in the expanding shell of V5668 Sgr, where \bevii\ is observed,  could  reduce substantially the
abundance of derived \bevii.

Following the recent detection  of \beviiii\   in the   outburst spectra of  Classical Novae,  we  started an observing program at ESO to target \bevii\ in  novae which at maximum reach magnitude  V $\le$ 9 mag. We report here on the observations of V6595 Sgr  by means of the high resolution UVES spectrograph \citep{Dekker2000} at the  Very Large Telescope (VLT).

\begin{figure}
\includegraphics[width=0.99\columnwidth]{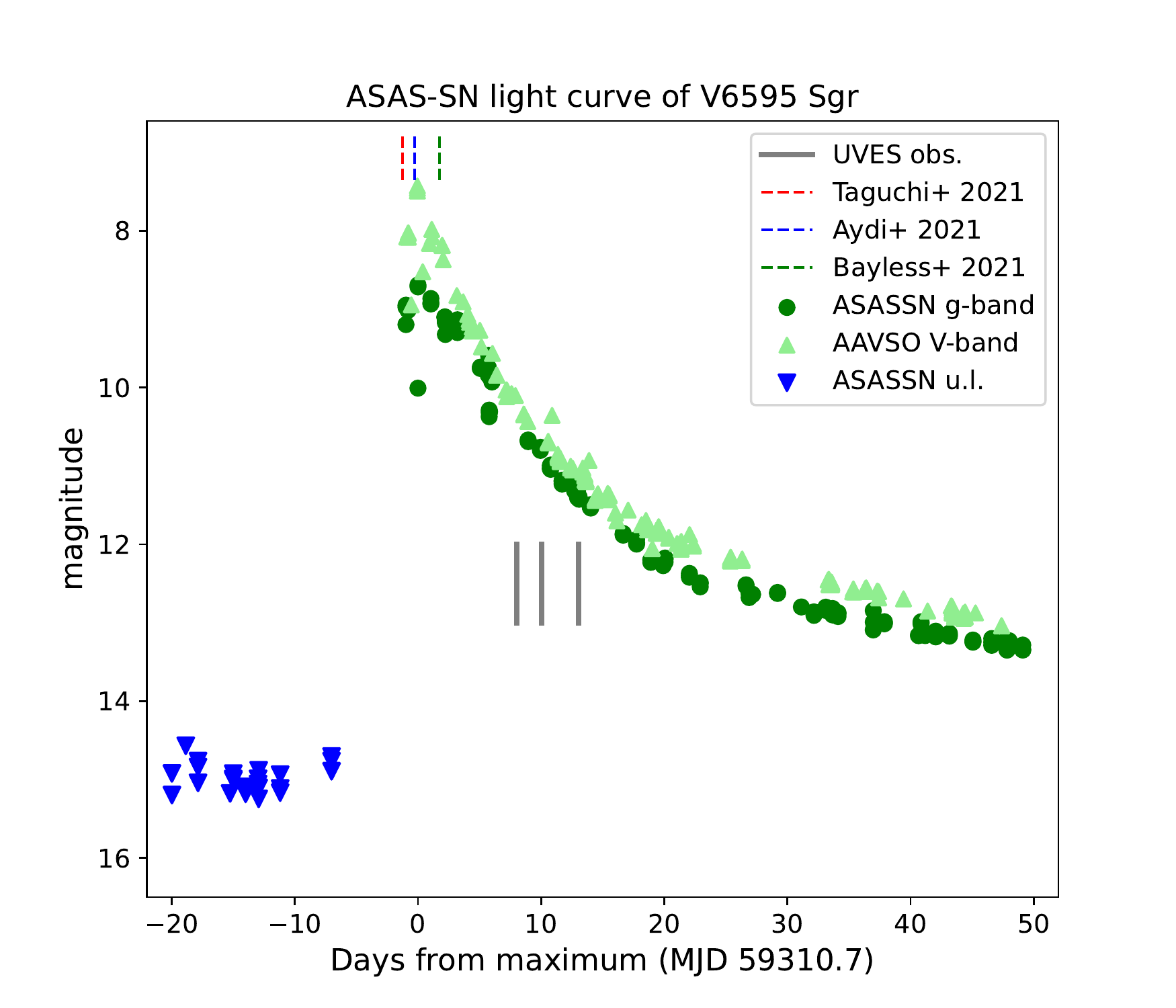}
\caption{The ASAS-SN \citep{Kochanek2017} g-band (green circles) and the AAVSO \citep{Kafka2021} V-band (light green triangles) light curves of V6595 Sgr. The blue triangles correspond to ASAS-SN upper limits. The black dashed lines mark the epochs of the VLT/UVES spectroscopic observations, presented in this work.}
\label{fig:1}
\end{figure}

\begin{figure*}
\includegraphics[width=19.0cm]{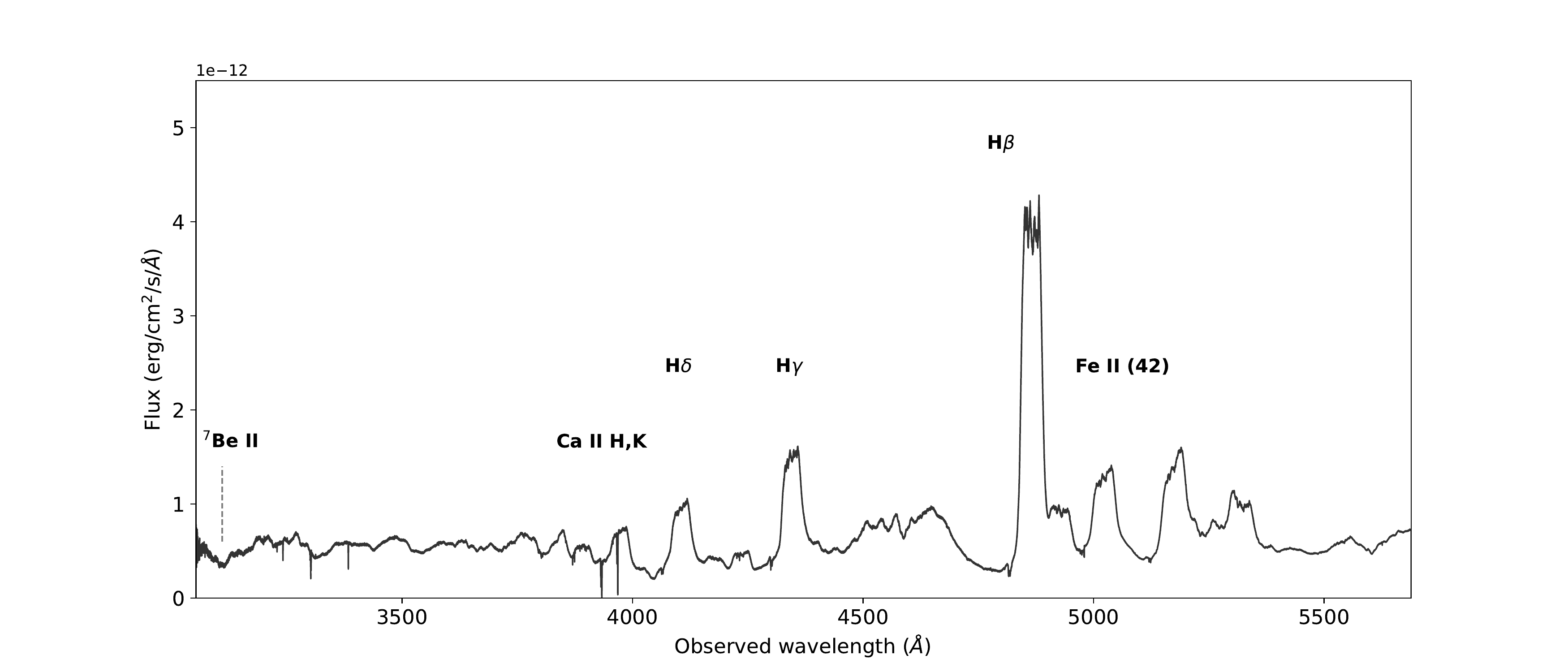}
\includegraphics[width=19.0cm]{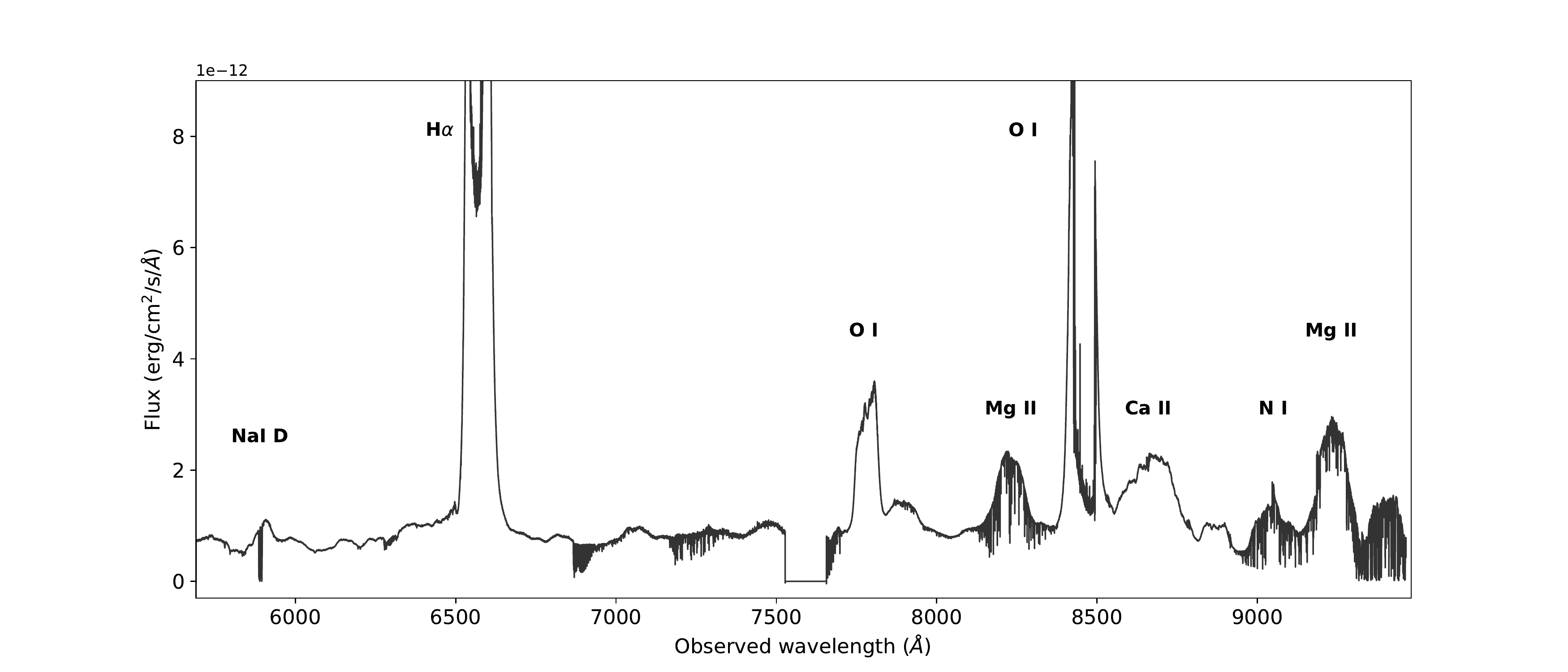}
\caption{The spectrum of V6595 Sgr obtained at VLT/UVES on 2021 April 16. The upper panel shows the blue wavelength region (305-570 nm) while in the lower panel the red region (570-950 nm) is displayed. Some of the main visible features, also used for the analysis presented in this work, are marked in both panels. The H$\alpha$ and O I 844.6 nm features are saturated. The spectrum has been corrected for the interstellar extinction, using the value of $A_V = 1.26$ mag, as reported in \citet{Bayless2021ATel14553....1B}.}\label{fig:1aa}
\end{figure*}

\begin{figure*}
\includegraphics[width=19.0cm]{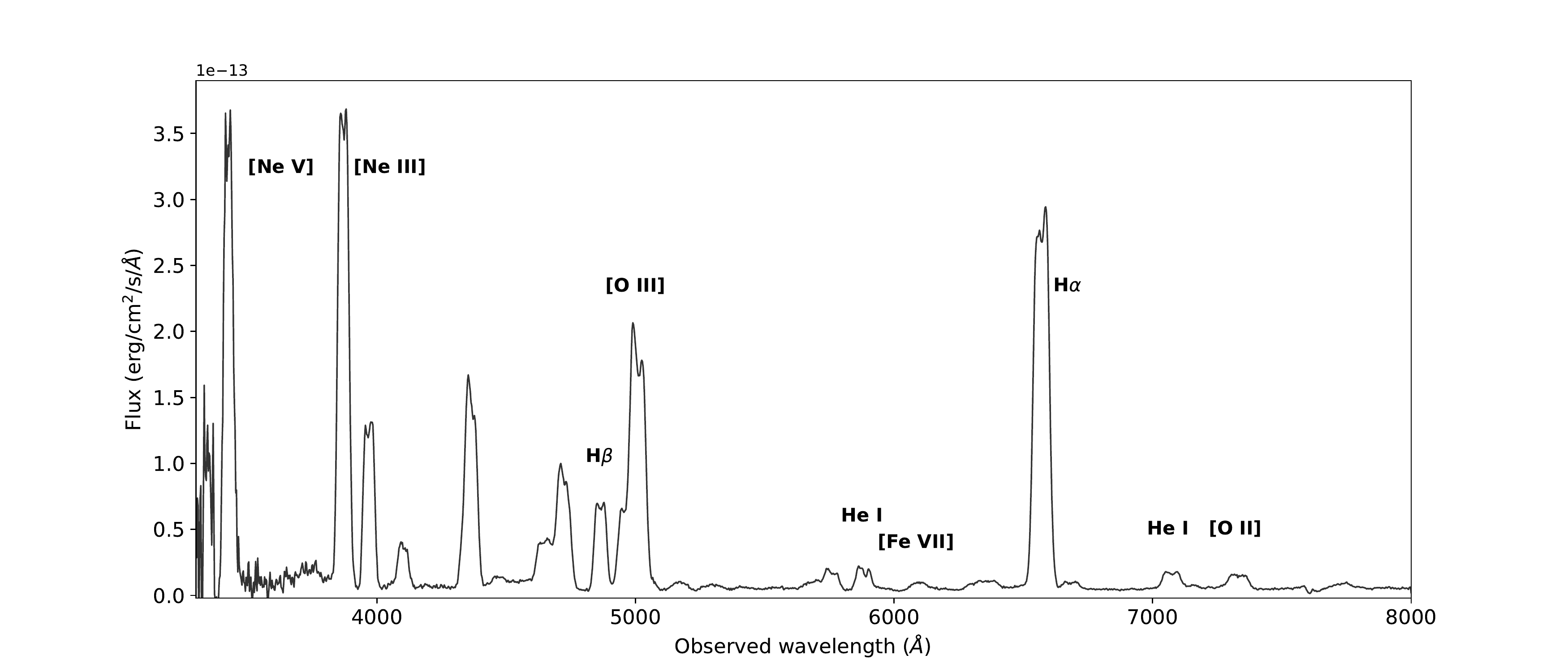}
\caption{  The TNG/DOLORES spectrum of V6595 Sgr obtained 91 days after the nova discovery. The spectrum shows prominent emission lines of Oxygen and Neon allowing us to classify the nova as a ONe type. The spectrum has been corrected for the interstelar extinction, using the color excess value of $E(B-V)$ = 0.65 mag \citep{Bayless2021ATel14553....1B}.  Note that the flux level at wavelengths $<$ 350 nm  is not very accurate, given the low transmission efficiency of the  camera.}\label{fig:1aaa}
\end{figure*}

\begin{figure}
\includegraphics[width=0.99\columnwidth]{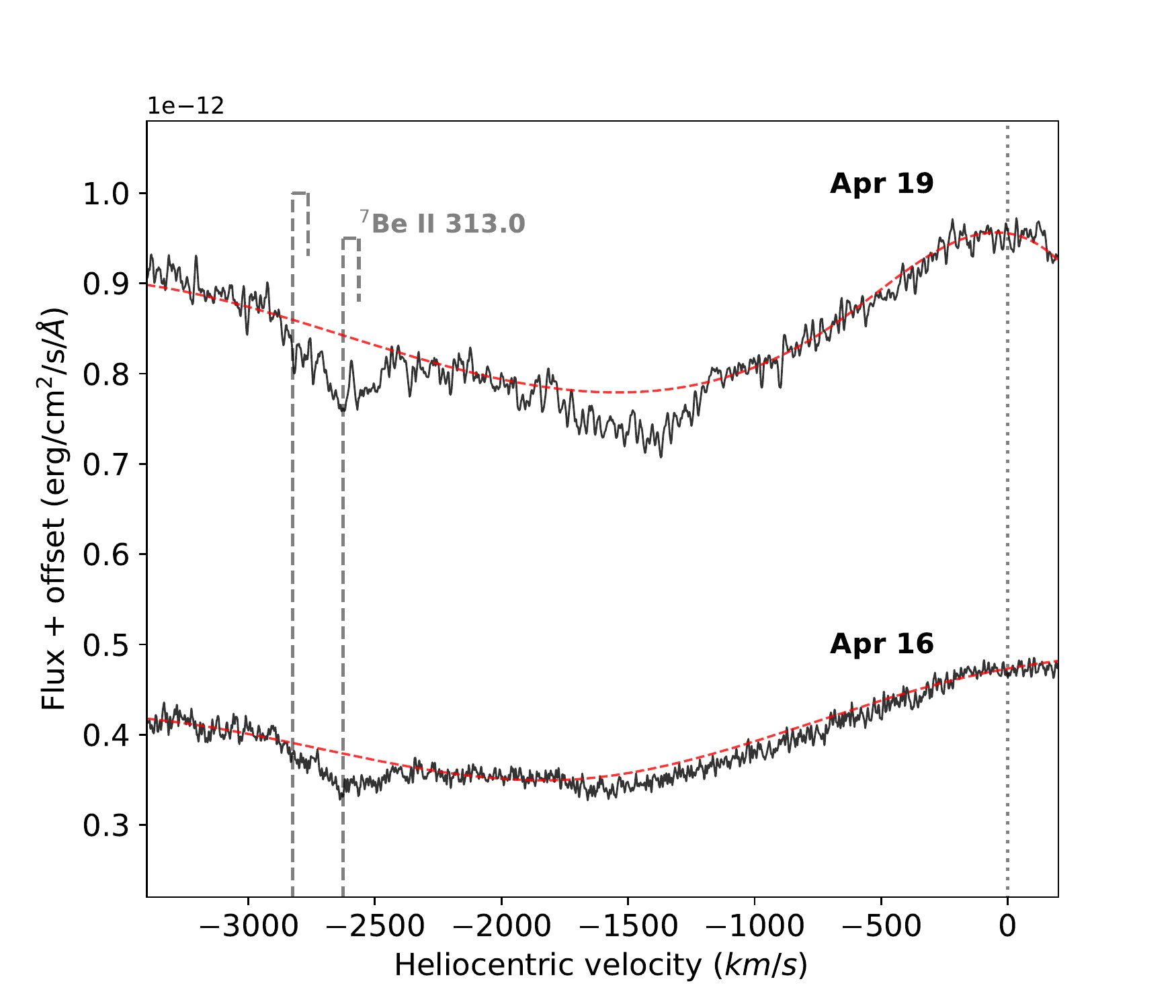}
\caption{ The $^7$Be II 313 nm feature in two  spectral epochs obtained with VLT/UVES.The spectra are slightly smoothed to increase the SNR. The dashed gray lines mark the position of the $^7$Be II 313.0 nm doublet at the blue-shifted heliocentric velocities  of  -2625  and   -2810 \kms and  the continuum is overlayed.  The zero of the scale  is set at \beviiii $\lambda 313.0583$ nm. \label{fig:1a}}
\end{figure}

\begin{table}\label{tab:11}
\caption{V6595 Sgr: basic data }
\begin{center}
%\scriptsize
\begin{tabular}{lr}
\hline
%\multicolumn{1}{c}{}& 
%\multicolumn{1}{c}{}&
%\multicolumn{1}{c}{}\\
\hline
Property & Value \\
\hline
Discovery & 2021 April 04.825 UT   \\
RA (J2000) & 17 58 16.7  \\
Dec (J2000) & -29 14 49.0 \\
Epoch max. & 2021 April 06.2 UT \\
g$_{max}$ & 8.69 mag  \\
t$_2$ & 8.9 days\\
t$_3$ & 15.3 days\\
$E(B-V)$ & 0.65 mag\\
$A$ & 9.49 $\pm$ 0.40 mag\\
$d_{MMRD}$ & 8.6 $\pm$ 0.9 kpc\\
%M$_v$ && \\
%d && \\
\hline
\hline
\end{tabular}
\end{center}
\end{table}

\section{V6595  Sgr   }

 V6595 Sgr, originally reported as PNV J17581670-2914490\footnote{http://tamkin1.eps.harvard.edu/unconf/followups/J17581670-2914490.html}, was discovered  on 2021 April 04.825 UT by Andrew Pearce at 8.4 mag. The nova was classified as a classical nova  by \citet{taguchi2021} and confirmed in follow-up spectra by \citet{Aydi2021ATel14533....1A}. The ASAS-SN \citep{Kochanek2017} light curve of V6595 Sgr is displayed in Fig \ref{fig:1}. The photometry shows that the nova is a very fast one, according to the historical classification by \citet{Payne1957}: it reached optical peak on 2021 April 06.2, i.e. 1.4 days after its discovery, and faded by two magnitudes in 8.9 days.  At the position reported in \citet{Ferreira2021}, there is a single source in the GAIA EDR3 catalog \citep{GAIADR3}, namely GAIA 4062339124163281920. The GAIA  magnitude  is $g$ = 18.18 mag, suggesting an amplitude of the outburst of $A$ = 9.49 $\pm$ 0.40 mag, with the uncertainty related to the different photometric systems used in  GAIA and ASAS-SN surveys\footnote{see e.g. the GAIA DR documentation release at the link: https://gea.esac.esa.int/archive/documentation/GDR2/Data\_processing/ \\
 chap\_cu5pho/sec\_cu5pho\_calibr/ssec\_cu5pho\_PhotTransf.html}. No parallax is reported for this source in the DR3. Given the observed magnitude at peak of $V$ = 7.6 mag (see Fig. \ref{fig:1}), the reddening $R_V$ = 2.01 mag, and the decay time of $t_2$ = 8.9 days, we can derive the absolute magnitude at the peak using the recent formulation of the maximum-magnitude and rate-decay (MMRD) relation given in \citet{DellaValleIzzo2020}: we obtain $M_V$ = -9.08 $\pm$ 0.22 mag, resulting in a nova distance  of $d_{MMRD}$ = 8.6 $\pm$ 0.9 kpc.    The estimate from the MMRD relation is in good agreement
with that estimated from the geometric distance
($M_V$= -8.89 $\pm $ 0.7) and photogeometric distance ($M_V$ = -9.03 $\pm $ 0.8)  \citep{bailer2021AJ....161..147B}. The basic data for the V6595 Sgr are summarized in Tab. \ref{tab:11}.
 
In  early spectra of V6595 Sgr \citet{Aydi2021ATel14533....1A} noted  P-Cygni profiles and emission lines of low contrast relative to the continuum but with exceptional breadth. The \nai\  D interstellar absorption shows multiple components at velocities  from 0 up to  -215 \kms, suggesting   high extinction towards the line of sight. From  Diffuse Interstellar Bands, \citet{Aydi2021ATel14533....1A}  derived a colour excess $E(B-V)$ = 0.65 mag,  and $A_v$ = 2 mag when  assuming the standard Galactic extinction law $R_V$ = 3.1. They also found this consistent with the reddening estimated from the Galactic reddening maps.  Observation taken  few  days later by \citet{Bayless2021ATel14553....1B}  revealed strong  emission lines,  while the  P-Cygni absorption  mostly disappeared. They also noted that the  emission lines were  exceptionally broad  with FWHM of 3500 \kms or greater. Fe II features were still present  along with C I, N I, and O I. The N I lines are unusually strong relative to  C I, which together with  the presence of  He I,  indicate a rapid spectral evolution. 

No X-ray emission has been detected by Swift/XRT   on 2021 April 5.86 \citep{Sokolovsky2021ATel14535....1S}. \citet{McCollum2021ATel14655....1M}  performed SED fitting of the available photometry for the progenitor object and with a  set of model atmospheres  found the best fit  for Teff = 3750 $\pm$ 150 K, and log g = 3.5  $\pm$ 0.3. The value of $A_V$ obtained from their best fit analysis is 1.26 $\pm$  0.22 mag. In the following, we will use this value to correct our spectral dataset for  Galactic extinction. 

Three spectra   for  V6595 Sgr  were obtained at VLT/UVES following the  peak brightness as soon as  possible due to the critical pandemic circumstances.  The UVES settings used were DIC1 346-564,  with ranges 305-388\,nm  in the blue  and  460-665\,nm in the red, and    DIC2 437-760 with ranges 360-480\,nm  in the blue,  and  600-800\,nm in the red.    The journal of the observations for the  nova is  provided in  Table \ref{tab:1}. The nominal resolving power   of the first two epochs was of  $R= \lambda /\delta \lambda  \approx 80,000$ for the blue arm and $\approx 120,000$ for the red arm. In the third observation the slit was set to 1.2 arcsec and the pixels were binned two by two to cope with the nova fading, degrading by   about half  the previous resolution.  Overlapping spectra were combined for each epoch to maximise the signal-to-noise ratio.
In Fig. \ref{fig:1aa}, we  show the spectrum of V6595 Sgr obtained at VLT/UVES on 2021 April 16.

\begin{table}\label{tab:1}
\caption{Journal of  UVES observations and  setting parameters. }
%\begin{center}
\scriptsize
\begin{tabular}{rrrrrrrr}
\hline
\hline
\multicolumn{1}{c}{{Date}} & 
\multicolumn{1}{c}{{Time}} & 
\multicolumn{1}{c}{{B,R slit}} & 
\multicolumn{1}{c}{Bin} & 
\multicolumn{1}{c}{346 nm} &
\multicolumn{1}{c}{437 nm} & 
\multicolumn{1}{c}{564 nm} &
\multicolumn{1}{c}{760 nm}  \\
\multicolumn{1}{c}{} & 
\multicolumn{1}{c}{} &
\multicolumn{1}{c}{arcsec}&
\multicolumn{1}{c}{pix} & 
\multicolumn{1}{c}{sec}& 
\multicolumn{1}{c}{sec}& 
\multicolumn{1}{c}{sec}&
\multicolumn{1}{c}{sec}\\
\hline
\hline
   2021-04-14  & 7:12 &0.4, 0.3 &1x1&  & 60  &  & 60 \\
    2021-04-14  & 7:25 &0.4, 0.3 &1x1& 300 &   &60 &  \\
2021-04-16& 8:04 &1.2, 1.2 &1x1 &   &700 &&700  \\
2021-04-16& 8:44 &1.2, 1.2 &1x1 &  2310 & &2310&  \\
      2021-04-19 &7:21 &1.2, 1.2 & 2x2  &  & 900 & & 900  \\
       2021-04-19 &6:11 &1.2, 1.2 & 2x2  & 3600 &  & 3600 &   \\
\hline
\hline
\end{tabular}
%\end{center}
\end{table}

 A spectrum of V6595 Sgr was also obtained on July 4, 2021, 91 days after  discovery, by using the 3.6m Telescopio Nazionale Galileo (TNG) equipped with the Device Optimized for the LOw RESolution (DOLORES) instrument in low-resolution spectroscopy mode. Multiple exposures were obtained using the LR-B grism to optimize the signal-to-noise at wavelengths lower  than 400 nm, and to avoid possible saturation of bright emission lines. The stacked spectrum covers the wavelength region between 330 and 800 nm and  is shown in Fig. \ref{fig:1aaa}. The spectrum shows typical emission lines observed in the nebular phase of classical novae such as [\oiii], [\oii] and [\nii] \citep{DellaValleIzzo2020}. In particular, we identified bright forbidden [\neiii]  and [\nev] lines, with [\neiii] 386.9 nm being the most bright line in the spectrum, see Fig. \ref{fig:1aaa}. This  suggests an over abundance of Neon in the  ejecta, which implies that V6595 is a ONe nova type \citep{Williams1991a}. We also note the presence of relatively faint high-ionization iron forbidden lines which are commonly observed during the super-soft phase of classical novae \citep{Ness2007,Schwarz2011}.    

\section{ \bevii\ Detection \& abundance}

\begin{figure}
\includegraphics[width=0.99\columnwidth]{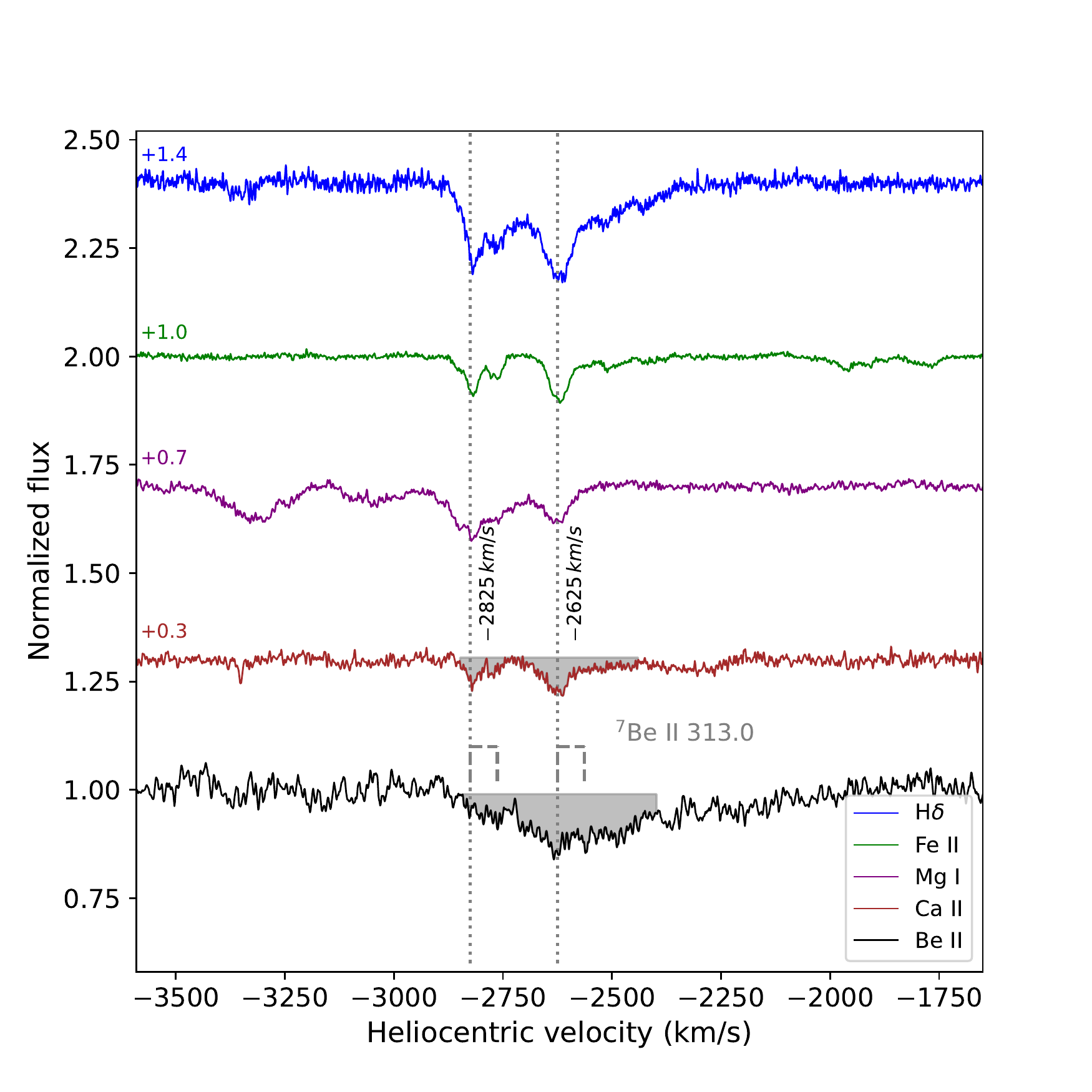}
\caption{   Normalized spectrum of 16 April  in the regions of  \beviiii,   \caii\ K, \mgi\ 383.829 nm, \feii\ 516.903 nm and H$\delta$  lines from bottom to top  respectively. The normalization has been performed by using the adjacent regions around the absorption. The narrow absorption features are marked with a dotted line while the shadowed area show the shallow absorption.}\label{fig:1b}
\end{figure}

\begin{figure}
\includegraphics[width=0.99\columnwidth]{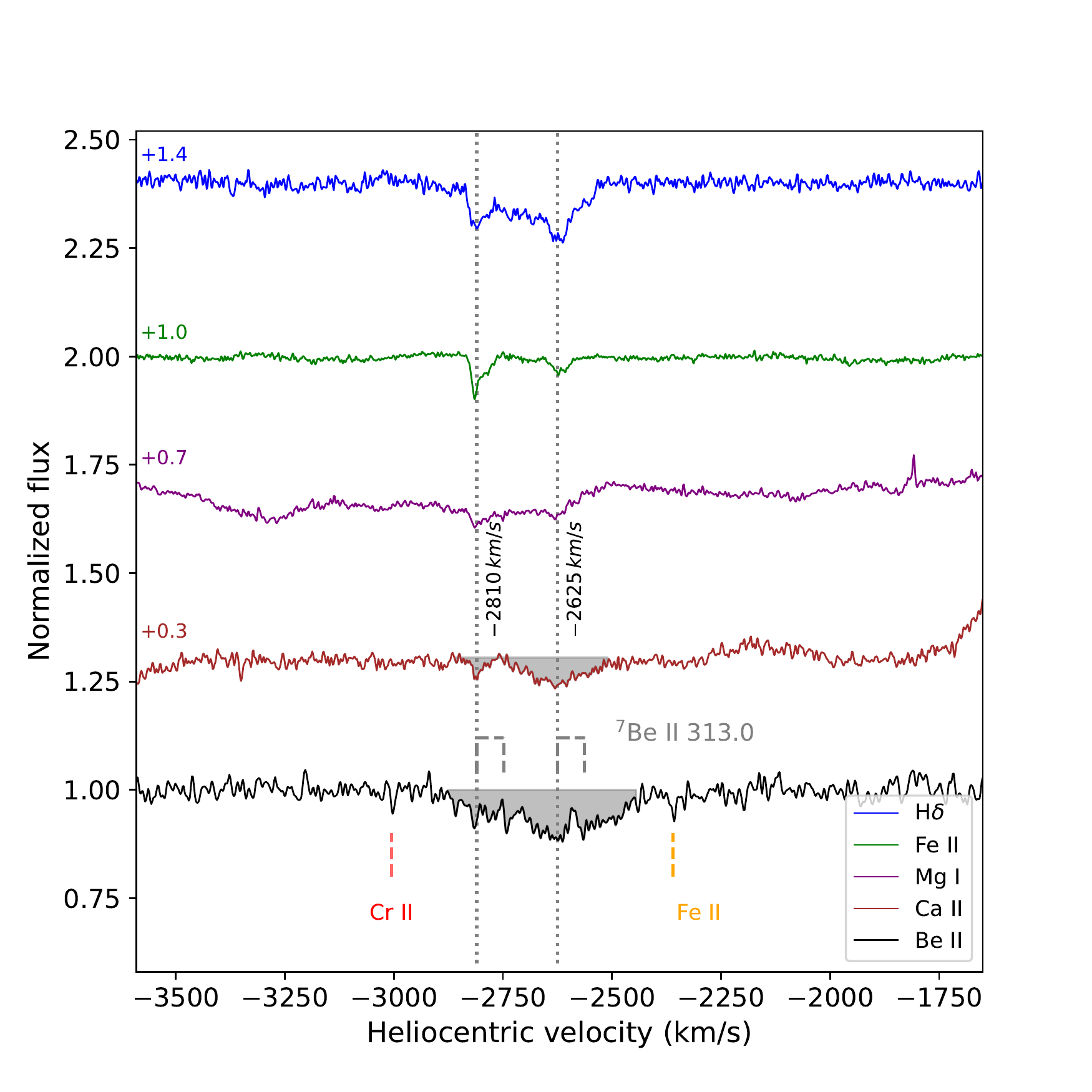}
\caption{  As in the previous figure but  for  the spectrum of 19 April. Note that in this spectrum the faint and sharp lines for the -2810 \kms components of  \crii\ 312.870 nm} and \feii\ 313.536 nm can be identified in the spectrum.\label{fig:1c}
\end{figure}

The three  UVES spectra covering  the Beryllium region in the three epochs  are shown in Fig. \ref{fig:1a}.
The outburst spectra  show absorption  in a range of radial velocities spanning from -2000 to -3000 \kms. 
Fig. \ref{fig:1b} and \ref{fig:1c} show the \bevii~spectra  of 16 and 19 April, respectively,  along with the portions with the  \caii\ K,  and H$\delta$  lines  on a common velocity scale and normalized with a local continuum around the  highlighted features. H$\delta$  is representative of the absorption  seen in all members of the Balmer Series. The \caii\ H cannot be used since it is contaminated with H$_\epsilon$ and for this nova  its high velocity components are blended with the Galactic interstellar components of  \caii\ K. The appearance of the absorption is clear in all  three  elements.
Superposed to a relatively shallow  absorption there are small but narrower structures   at velocities of -2630 and  -2815   \kms whose  positions  are  marked on the figure.
 In all  three spectra there is a hint of  features which could be possibly ascribed to other lines of the  \beviiii~ doublet at a separation of 62 \kms.  We preferred not to average the 3 spectra since, although taken rather close in time, a small velocity drift of the components  is always possible. The narrow features are hidden in the noise,  but their presence in all the three spectra make us confident they are real.
 We can also detect very weak lines of \crii\ 312.870 nm and \feii\ 313.536 nm at -2810   \kms.  Since these weak lines are the strongest lines in the multiplets they show that the contamination in the region is minimal and   both \crii\ and \feii\  cannot be responsible for the observed absorption.
 It is also possible that the  absorption could be in the form of a P-Cygni with an \bevii~ emission centered at rest wavelength and a large absorption shortwards. An emission in \beviiii~ has been detected in  the oxygen poor  Nova V838 Her by \citet{Selvelli2018}. However, here we focus on the absorption for which there is correspondence with other elements.

 Relatively weak components make the detection more problematic but  provide more reliable abundance estimation since saturation effects can be ignored.  \citet{shore2020}   showed that with high abundances the \beviiii~  lines are strongly saturated and  cannot be used. 
The  \bevii\ abundance  are generally estimated taking  Ca,  which is not a nova product, as reference.  \mgii, if available, could  also be used for novae  of the ONeMg type \citep{casanova2016}.  

We do not detect   doubly ionized ions, whose transitions are  generally  characterized by  higher ionization potentials.  On the other hand we see the    presence of   neutral species such as \mgi\ 383.8, 383.2, 379.8 e 379.6 nm.
Therefore we assume that  singly ionized ions  of \caii\ and \beviiii\ are representing  the main ionization  stage  in the expanding shell.  This assumption has been criticised by \citet{chugai2020} in the nova V5668 Sgr and will be addressed  in detail in the next section.
  
Since we cannot resolve the \beviiii\ doublet for the shallow absorption  we consider   the equivalent width (EW) of the sum of the \beviiii\, $\lambda313.0583 + \lambda313.1228$  doublet and  compare it with the \caii~  K line at 393.366 nm \citep{Tajitsu2015,Molaro2016}. We have 

\begin{eqnarray}\label{eq:1}
  \frac{N(\mbox{\beviiii})}{N(\mbox{\caii})}
  & = &   2.164   \times \frac{EW(\mbox{\beviiii\,, Doublet})}{EW(\mbox{\caii\,, K})} 
\end{eqnarray}

\noindent
 where $log (gf)$ of -0.178 and -0.479 for the \beviiii\ doublet, and    +0.135 for that of \caii\,K  line are adopted.
The equivalent width  of the whole region spanned by the \beviiii\ components  as measured in the spectrum of 16 April is of  274 $\pm$ 15   m\AA\ with  the main  uncertainty  coming from  the continuum placement.   The equivalent width  of the \caii\ 393.3 nm  line is of 189 $\pm$ 8  m\AA\  providing  a ratio of   EW(\beviiii)/EW(\caii) $\approx$ 1.45 $\pm$ 0.15. The same  components  as measured in the spectrum of 19 April are  of $\approx$ 279 $\pm$ 13   m\AA\ and  165 $\pm$ 7  m\AA\  providing  a ratio of  1.69 $\pm$ 0.15.  We therefore adopt an average ratio of 1.57 $\pm 0.15 $.
Using  Eq. \ref{eq:1},  we obtain   N(\beviiii)/N(\caii) = 3.4 $\pm 0.15 $. Since \bevii\  is unstable with  a  half life time decay of 53.22  days, at the epoch of our measurement, 15 days from discovery, the    abundance   had  reduced by a factor of 0.745.  The  solar abundance  of  calcium is  N(Ca)/N(H) = 2.19 $ \pm 0.30 \cdot 10^{-6}$  \citep{lodders2019arXiv191200844L}. As a consequence,  we obtain an abundance of   N(\bevii)/N(H) %=1.57x2.164x2.18/0.754=
= 9.8 $\cdot 10^{-6}$.  Should Calcium abundance   be different from  solar,   the final N(\bevii) /N(H) would change  accordingly. For instance, by extrapolating  the Galactic metalliciy gradient 
of \citet{lemasle2018A&A...618A.160L} towards the Galactic Center  we should expect a metallicity of [Ca/H]$\approx$  +0.3,  which would  imply N(\bevii)/N(H) 
= 2.0  $\cdot 10^{-5}$, namely a factor 2 higher  than what reported in Table \ref{tab:3}.

\subsection{Lithium}

\begin{figure}
\includegraphics[width=0.99\columnwidth]{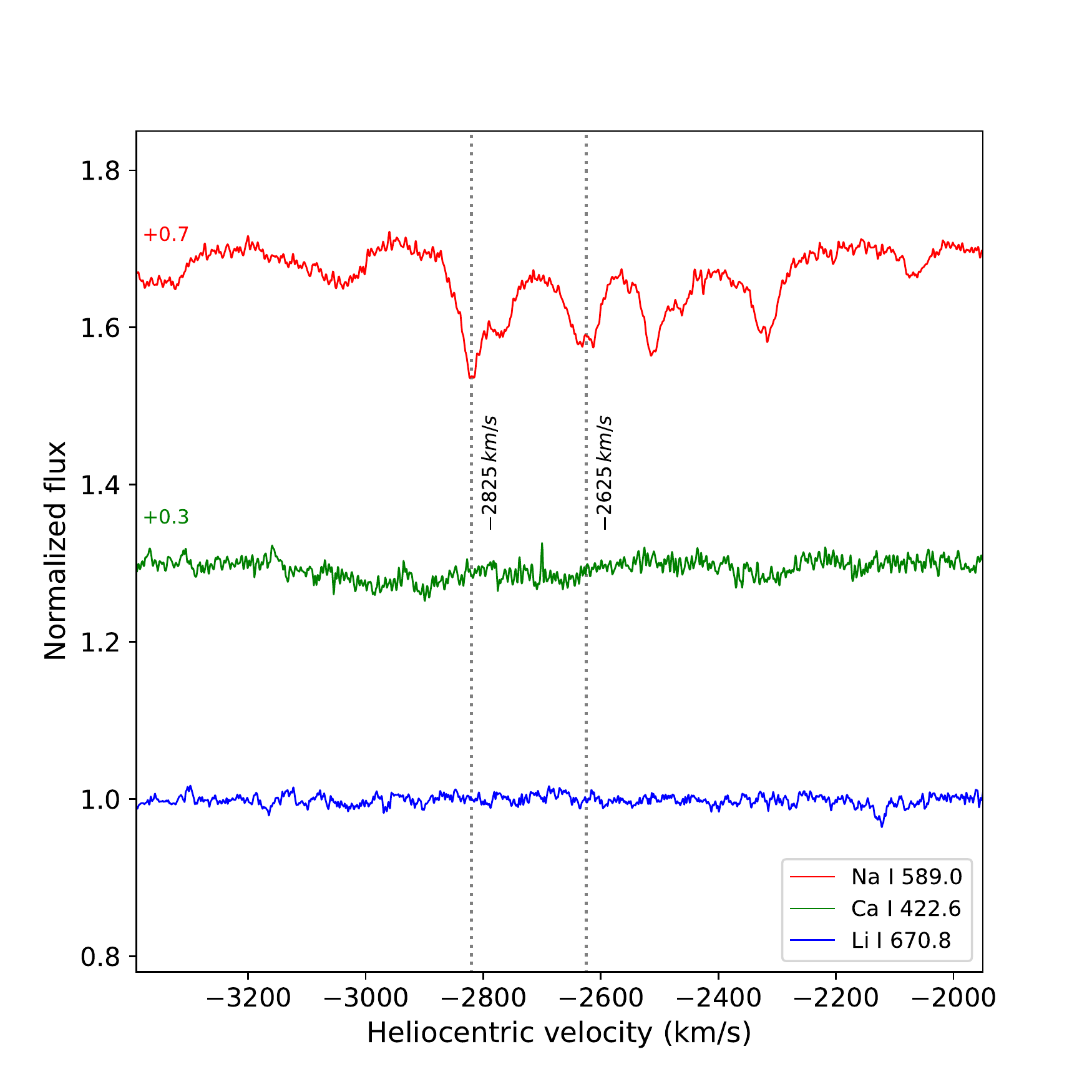}
\caption{  Spectra of 16 April in the regions of \nai\, 589.0 nm \cai\ 422.6 nm and \liviii\ 670.8 nm. While blue-shifted expanding absorptions are clearly detected for the Na I 589.0 nm line, there is no evidence of \cai\ and \liviii\ at the same expanding velocities. The absorptions of the D1 line are also visible redwards the D2 line. }
\label{fig:1d}
\end{figure}

 In Fig. \ref{fig:1d} the spectrum of 16 April is shown in the region of \liviii\ 670.8 nm together with the D2 line of \nai\ 589.0 nm and \cai\ 422.6 nm. Absorptions are detected in correspondence of the D2 and D1   lines with expasion velocities of -2625 and -2825 \kms but not in correspondence of the  \cai\ and \lii\ lines.  The same is found for the spectra of the other two epochs of 14 and 19 April. Thus, there is no  evidence of the resonance \liviii\ doublet at 670.8 nm in the spectra of V 6596 Sgr  as in most of the novae. So far  \liviii\ has been detected only in V1369 \citep{Izzo2015} and, as a trace element, in a very early epoch of    V906 Car  \citep{molaro2020MNRAS.492.4975M}.
The  detection in V1369 was determined  on days 7 and 13 when  only a small amount of \bevii\ had decayed into \livii\  and implies that the TNR started much  earlier than the explosion \citep{Izzo2015}. A claim for a Li detection was made by \citet{DellaValle2002} in V382 Vel but  \citet{Shore2003AJ....125.1507S} suggested that the feature could be neutral nitrogen. The general absence  of neutral $^7$Li on a time scale longer than the \bevii\ decay can be explained if  \bevii\ decays through the   capture of an internal K-electron.  In this case  the end product is  an  ionized lithium with transition lines in the X-ray \citep{Molaro2016}.

%\begin{figure}
%\includegraphics[width=0.99\columnwidth]{epoch2_MgI.pdf}
%\caption{ \mgi\ 383.829 nm.}\label{fig:1c}
%\end{figure}

%\begin{table}
%\caption{ Cloudy results for a  model with a Black Body with T=5000 K and  r=10$^{13}$ .}
%\label{tab:2b}
%\begin{center}
%\begin{tabular}{lrcc}
%\hline
%\hline
%Element  &    & Ion &    \\
% &  I & II &   III   \\
%\hline
%\hline
%Hydrogen  &  20.518 &17.322 &\\
%Helium    & 19.518 & 13.777&\\
%Lithium   & 6.216 &11.828&\\
%Beryllium & 8.051  &9.932&\\ 
%%Boron     &  7.602 &11.308&\\
%%Carbon    & 15.870 &16.865& \\
%%Nitrogen  & 16.448 &12.784&\\
%%Oxygen    & 17.208 &13.904&\\
%Sodium    & 10.523 &14.849&\\
%Magnesium & 12.626 &16.058& 11.084\\
%Calcium   &  9.211 &14.865& 13.345\\
%\hline
%\end{tabular}
%\end{center}
%\end{table}

\section{Discussion}

\subsection{Over-ionization?}
  The \beii~  abundances measured in the novae where it has been detected so far are  high, see Table \ref{tab:3}, and in particular are higher than  what the nova models  predict. 
 Recent reappraisal of the theoretical  \bevii~ maximum which can be synthesized in novae is of X(\bevii)/X(H) $\approx$ 2 $\cdot 10^{-5}$ as mass fraction , or N(\bevii)/N(H) $\approx 3  \cdot 10^{-6}$ in atomic fraction, for CO and relatively massive white dwarfs \citep{starrfield2020ApJ...895...70S, Deenissenkov2021MNRAS.501L..33D, chugai2020}.   \bevii~ abundances are  generally derived assuming \beviiii\ and  \caii~are  both in their dominant ionization stage.
 \citet{chugai2020}  suggested  that the disagreement  between theoretical predictions and observations could be resolved if   significant  over-ionization  is present in the expanding components. Since the ionisation potential of \beii~ and \caii~ are  18.21 and 11.87 eV, respectively, the radiation from  a hot  photosphere  could over-ionize \caii~ and reduce the \bevii~ abundance derived from observations.
 
  In nova  V838 Her \citet{Selvelli2018} obtained  the   \bevii\  abundance by using \mgii~ as a reference element. Mg  has a second  ionisation potential  of 15 eV which is  closer to \beii\ than \caii\ and the two ions should behave similarly. They obtained   N(\bevii)/N(H)  $\approx 2 \cdot 10^{-5} $, a value  which is similar to the value measured in  the other novae where \caii~ was used and listed  in Table \ref{tab:3}.    \citet{Selvelli2018}  also derived  consistent abundances  by using  \hei\ and \hi\  emission lines. A  consistency which would  not be expected  in the presence of significant over-ionization.

\citet{chugai2020} analysed in detail  the case of Nova V5668 Sgr  studied by \citet{Molaro2016} and \citet{Tajitsu2016}, and  concentrate on the component  at  velocity  -1175 \kms~  shown at day 58. 
No doubly ionized states of Fe-peak elements with second ionization potentials intermediate  those of \bevii\ (18.21 eV) and \caii\  (11.87 eV) are observed in correspondence of this high velocity component  of nova V5668 Sgr \citep{Molaro2016,Tajitsu2016}. This was one  major justification  used  in \citet{Tajitsu2015,Molaro2016} to consider  \bevii\  and  \caii~ in the main ionization state  in  V5668. On the other hand, 
 hints of the neutral species of  \cai~ and \nai, see Fig 7 of \citet{Molaro2016}, are reported. For any combination  of physical parameters it is  not possible to have  \caiii~ and  \cai\    in the same slab of material.

\citet{chugai2020}  argue for the presence of    partial covering   and of  multiple components. However,  they obtain a    \beii/\caii\  ratio of   36.6 which is  close  to the   value of 31.9  derived by \citet{Molaro2016} for this component. We note that  the measured   \beii/\caii = 31.9    implies a X(\bevii)/X(H)   mass fraction of 1.2 $\cdot 10^{-3} $  and not of 9 $\cdot 10^{-5} $ as reported by  \citet{chugai2020}.  Thus, a correction of two orders of magnitude is required to reconcile these abundances with the theoretical \bevii~ production. 
The pseudo  photosphere of the nova in the \citet{chugai2020}'s   model is   approximated with a black body with a  temperature of $\approx$  15000  K, and a radius of  10$^{12}$  cm,    for a  total luminosity of $\sim$3.6 $\cdot$ 10$^{37}$ erg/sec. The  high velocity component  showing \beviiii\  is located at a distance of 6x10$^{14}$  cm which is the distance reached after 58 days when moving with a velocity of  v$_{exp}$= -1175 \kms,  and it has  a thickness which is 10$\%$ of its radius. \citet{chugai2020} assume  a mass of 10$^{-5}$ M$_{\odot}$ for this component which with the assumed volume results into a gas density  of $\log (n_H) = \log (n_e)$ = 7.6 cm$^{-3}$. This shell
feels the  radiation coming from the nova pseudo-photosphere  and  \citet{chugai2020} estimate  a   \caiii/\caii $\approx$  10 in the   expanding shell with velocity of -1175 \kms. This  means that the \caii\  is not the main stage and the \bevii~ abundance previously estimated should be decreased by one order of magnitude.

We have used version 17.03 of the photoionization code Cloudy  \citep{ferland2017}  to probe photoionization conditions within the absorption component. In a first model, Model A, we adopted the same model of \citet{chugai2020} with a  temperature of the nova photosphere of 15000 K. The results, reported in  Table \ref{tab:2},   show that  \caii\ is   indeed much more ionized than \beii\ and even for a  larger fraction than that estimated by \citet{chugai2020}. However,  the question is whether the model provides a realistic description of nova photosphere and of the expanding shell were \bevii\ is observed.
%For instance, the radius of the expanding photosphere is assumed of $10^{12}$ cm  which, after 58 days,
%implies a mean  expansion velocity of 2 \kms. This  is likely too low   for the expansion of the bulk of the nova ejecta \citep{Gehrz1998,williams2013AJ....146...55W}. The radius of the photospheric envelope should be probably larger  and consequently its  temperature  lower to preserve  total luminosity. 
 The  gas  density in the expanding shell  is probably  higher than what assumed.  \citet{harvey2018A&A...611A...3H} at day 141  measure an electron density  of $\log (n_H)$ = 9 cm$^{-3}$ which is higher than the 7.6  assumed by \citet{chugai2020} 80 days earlier. Moreover, the formation of dust requires shielding from the WD's radiation which is possible if the gas density is greater than $\log (n_H)$ = 9-10 cm$^{-3}$ \citep{gehrz1987PNAS...84.6961G}.
 \citet{Gehrz2018}  found  evidence that dust condensation in V5668 Sgr commenced  82 days after outburst and  estimated an envelope  temperature of ~1090 K. \citet{Muztaba2020E&ES..537a2004M} modeled with Cloudy the spectrum of V5668 Sgr on 
 June 12, 2015, or day 89, with the nova in the dust production    phase and matching fairly well the emission spectrum with a temperature for the nova envelope of 6509 K. At day 107 \citealp{banerjee2016MNRAS.455L.109B} modelled the CO  emission line with a temperature of 4000 $\pm$ 300 K.
 
\begin{figure}
\includegraphics[width=0.99\columnwidth]{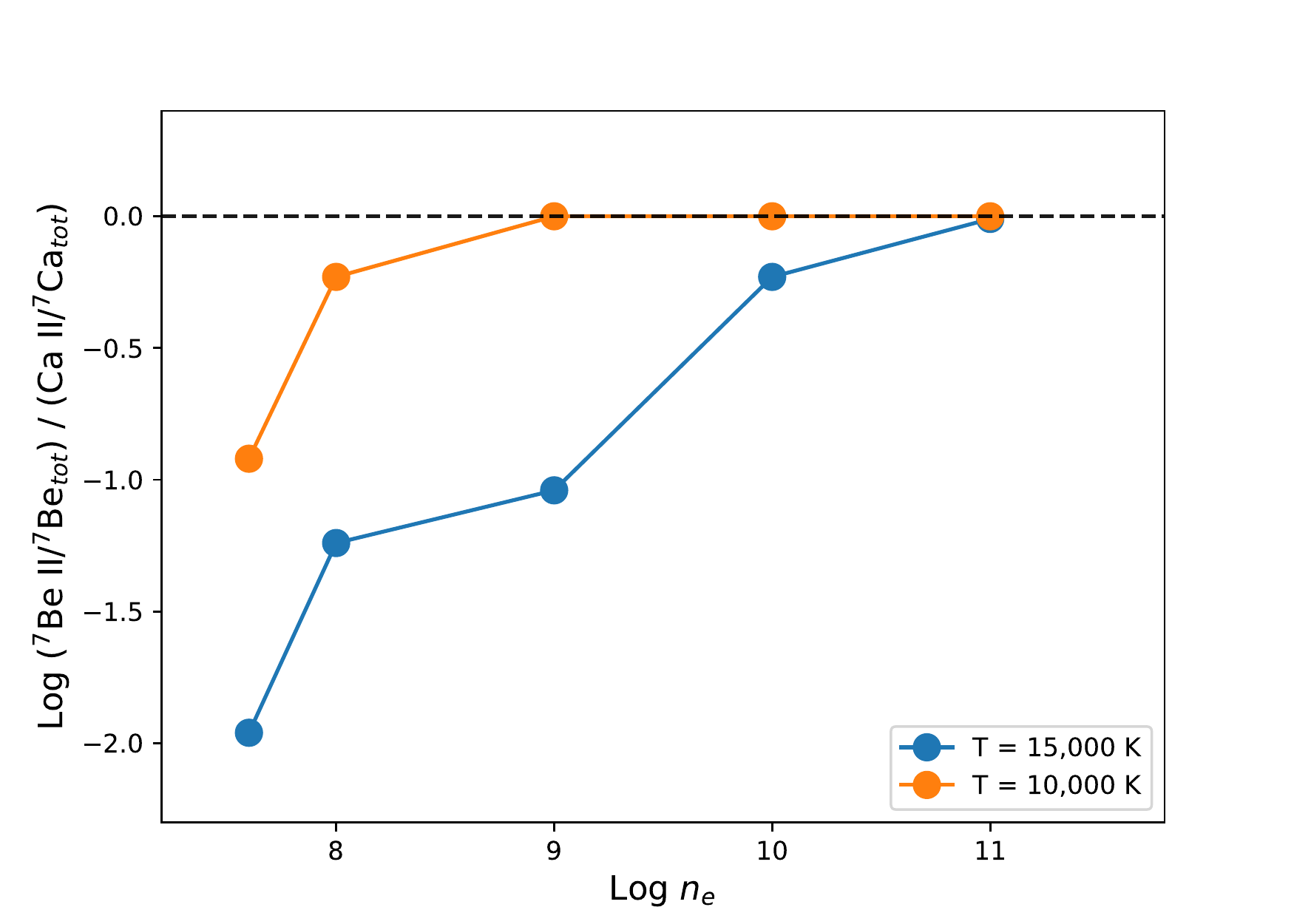}
\caption{   Relative mean ionization of \beviiii\ and \caii\, namely $\log (BeII/Be) / (CaII/Ca)$ for different values of gas densities and for two temperatures of the pseudo photosphere.}
\label{fig:overi}
\end{figure}

 We thus constructed a new  model, Model B,   with a temperature of T=15000 K for the nova envelope but with  a gas  density in the clump of $\log (N_H)$  = 10 cm$^{-3}$. The higher density is obtained reducing the thickness of the expanding shell and the filling factor. We have also constructed an intermediate model,
 Model C, with a black body temperature of 10000 K and a   gas  density  of  
 $\log (n_H)$ = 9  cm$^{-3}$. The mean ionizations for the relevant elements are listed in   Table \ref{tab:2}, In both models B and C   the  \caii~ and \beii~ are found in their main ionization stages.  The relative ionization fractions for \bevii\ and \caii\ for the two black body temperatures  for a wider range of gas densities are shown in Fig. \ref{fig:overi}. Thus, overionization is possible only for very low gas densities and rather hot pseudo-photosphere.

We emphasize that measured \bevii\ abundances in V5668 Sgr   obtained in components with different expanding velocities and also at different epochs show a consistency  \citep{Molaro2016}. For instance, at day 58 the component with v$_{exp}$ =-1175 \kms~ shows \beii/\caii = 31.9  while the component expanding with a velocity of -1500 \kms~  at day 82  shows  \beii/\caii =17.7 . Interestingly, these  two ratios become  69   and 53, respectively,   when the \bevii~  decaying factors are considered. We consider quite unlikely that the physical conditions  producing over-ionization  could remain   similar in two  shells expanding with different velocities and measured in  two   epochs separated  by a time interval of  24 days.

A satisfactory  physical model is presently not available  due to the  poor understanding  of the nova ejecta. However, for the arguments  provided here,  we consider unlikely that the disagreement between theoretical predictions by nova models and  \bevii~ abundances derived from observations could be ascribed to the presence of over-ionization in the expanding shells.

\begin{table*}
\caption{  Mean ionization for relevant elements obtained with Cloudy  for different conditions of the photosphere and for the expanding component. The logarithmic  fraction of an ionization stage over total is given. Model A is the model considered in  \citet{chugai2020}, with the photosphere of  a black body with temperature of  T=15000 K,radius    r=10$^{12}$ cm and  total luminosity of 38.3. The expanding shell with \bevii\ has a gas  density of $\log (n_H) = \log (n_H)$ = 7.6 cm$^{-3}$. Model B is as Model A but with  a gas  density in the clump of $\log (n_H) = \log (n_H)$ = 10 cm$^{-3}$. Model C is intermediate with  a black body temperature of   T=10000 K and  a gas density in the clump of $\log (n_H)$ = 9 cm$^{-3}$.}
\label{tab:2}
\begin{center}
\begin{tabular}{lrrrrrrrrrrr}
\hline
\hline
\multicolumn{1}{c}{El} &
\multicolumn{3}{c}{Model A} &
\multicolumn{4}{c}{Model B} &
\multicolumn{4}{c}{Model C} \\
 Ion&  I & II &   III  && I & II & III && I & II &III \\
\hline
\hline
H  &-0.71 &-0.09 & &|&  -0.03&-1.15& &|& -0.02&-1.40&\\       
He  & -0.01& -1.81& &|  & 0.00&-5.18& &| &0.00&-9.01&\\
Li   &-5.21 &0.00 & &|&  -3.83&0.00&&|&-4.68&0.00&\\
Be & -4.22& -0.28& -0.32&|&  -2.82&0.00&-4.55&|&-3.30&00&-762 \\
%Carbon    & 13.661 &17.829 &14.060 \\ 
%Nitrogen  & 16.708 &17.263 &11.746 \\ 
%Oxygen    & 18.045 &17.379 &10.127 \\ 
Na   &-3.82 &0.00 & -12.14 &|&  -2.54&0.00&-12.50&|&-3.23&0.00&-19.20 \\
Mg & -4.40& -0.31&-0.30 &|& -2.77&0.00&-2.85&|&-3.45&0.00&-3.11\\ 
Ca  &-7.82 &-2.24 &0.00 &|& -4.42&-0.23&-0.39&|&-5.05&-0.01&-1.55 \\
Cr & -5.79& -0.38&-0.23 &|&-4.14&0.00&-2.41&|&-4.56&0.00&-3.25 \\
Fe & -5.45&-0.35 &-0.26 &|&  -4.27&0.00&-2.84&|&-4.47&0.00&-3.64\\
  
\hline
\end{tabular}
\end{center}
\end{table*}

\subsection{\bevii\ nova yields}

\begin{figure}
\includegraphics[width=0.99\columnwidth]{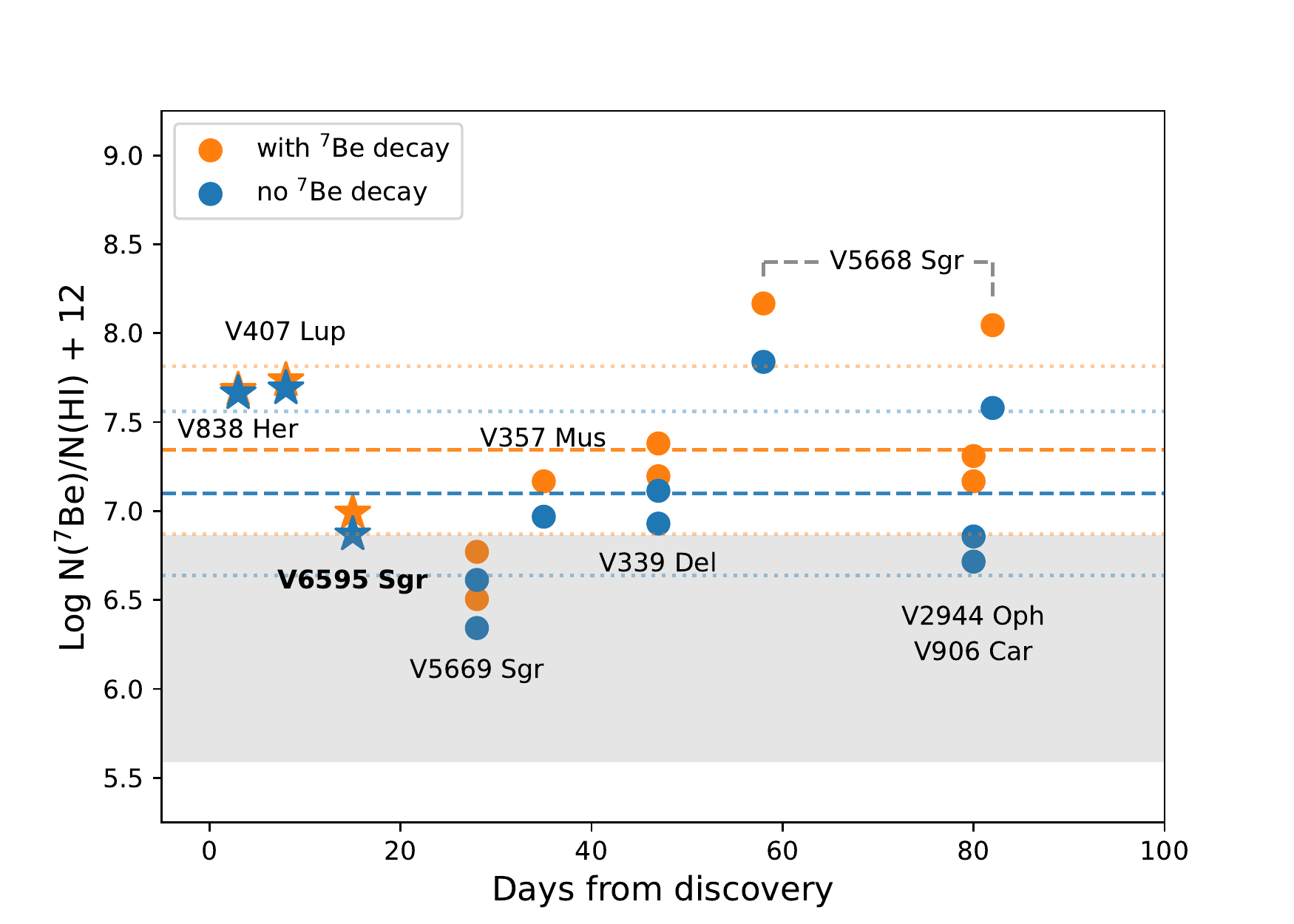}
\caption{   \bevii\  abundances as a function of the day of observation with, in orange, and without correction for the \bevii\  decay, in blue. The two
ONe novae are shown with a star while the 
CO novae with a circle.  The dashed lines mark the average value for both
estimates. $\log N($\bevii$/N(HI) +12 = 7.34 \pm 0.47 $ when \bevii\  decay is
taken into account, and $ \log N($\bevii$) /N(HI) +12 = 7.10 \pm 0.46 $, when it is not. The gray band shows the range of \bevii\ yields  for  WD of different masses according to \citet{starrfield2020ApJ...895...70S}.
With this scale the meteoritic    abundance is of A(\livii) = 3.3. The novae show a
mean overabundance of 4 orders of magnitude \citep{lodders2019arXiv191200844L}.}
\label{fig:abundances}
\end{figure}

The presence of \bevii\ in a high velocity component in the outburst
spectra of V6595 Sgr adds another object to the small sample of
novae where \bevii\  has been searched and found.  The A(\bevii) = $log N(^7Be)/N(HI) +12  $ abundances of these novae are
listed in Tab \ref{tab:3} and shown in Fig \ref{fig:abundances}. So far \bevii\  has been found
in all the novae where it has been searched off but  V612 Sct   which has been suggested could be peculiar \citep{mason2020A&A...635A.115M}. By means of a low resolution spectrum of V6595 Sgr
taken 91 days after the discovery which shows prominent lines
of forbidden \neiii\ and [Ne v] lines, we classified the nova as of
ONe type \citep{Williams1991a}. V6595 Sgr is together with V407
Lup a second nova of ONe type where \bevii\  has been detected.
Nova V838 Her is a third one but with some peculiarities being
deficient in oxygen \citep{Matheson1993ApJ...418L..29M}. The ONe novae show
quite different \bevii\  abundances with V407 Lup being about six
times higher than V6595 Sgr. The synthesis of \bevii\  is believed to
occur via  $^3$He($\alpha, \gamma $)\bevii\  in both types of CO and ONe novae but is
expected to be one order of magnitude higher in CO novae due
to the different time-scales of the TNR in the two types of novae
\citep{Jose1998}. The  ONe show \bevii\  abundances
which overlap those of the CO novae. While V6595 Sgr shows
among the lowest  abundance of the sample, V407 Lup is slightly higher   and V838 Her is  aligned with the mean
value of the CO novae. This supports the evidence that the \bevii\  abundances is
uncorrelated from the nova type. We note that  in the ONe type  the \bevii\ detection was made in the early phases when the nova pseudo-photosphere is  hotter. Thus, overionization is more likely to occur in fast novae,  and   the  \bevii\ of ONe could be  lower than  the CO type as expected from the TNR theory. 

The nine \bevii\  measurements show a  scatter which
exceed the observational errors and is likely real. The mean value is of A(\bevii) = 7.34 $\pm 0.47 $
when\bevii\  decay is taken into account, which is about 4 orders
of magnitude over the the meteoritic value \citep{lodders2019arXiv191200844L}.
The role of novae as \livii\  producers has been considered in a
framework of a detailed model of the chemical evolution of
the Milky Way \citep{Cescutti2019,Romano2021arXiv210611614R}.
\citet{Cescutti2019}  showed that novae account well for
the observed increase of Li abundance with metallicity in the
Galactic thin disk and also for the relative flatness observed
in the thick disk. In fact, the thick disk evolves on a timescale
which is shorter than the typical timescale for the production
of substantial\livii\  by novae. \citet{Cescutti2019} left the
nova yields as a free parameter and found that in order to match
the \livii\  behaviour with metallicity and the present abundance
a \livii\  production of  3$ \cdot 10^{-9} M\sun$   per nova event is required.
This is consistent with the measured mean yields of \livii/H of 1.5
$\cdot 10^{-4}$ in mass, and an average ejecta of 3 $\cdot 10^{-5} M\sun$. A constant nova
rate of $\approx$  20 yr$^{-1}$  during the Galactic life with these
yields is able to synthesize  $\approx$ 600 M$\sun$, i.e. about 60\% of the total
\livii\  estimated for the whole Galaxy, with the rest shared by Big
Bang, $\approx$ 250 M$\sun$, and Galactic spallation nucleosynthesis, $\approx$ 100 M$\sun$.

\citet{molaro2020MNRAS.492.4975M}  suggested that  a higher $^3$He  in the donor star could result into a higher \bevii\ since it  is produced trough the \iiihe($\alpha,\gamma$)\bevii\  channel.  \citet{Deenissenkov2021MNRAS.501L..33D} showed that a increased abundance of $^3$He in the accreted material leads to a decrease of the peak temperature, and therefore  a reduced production of \bevii. They  suggested instead that an enhanced abundance of $^4$He could favour the \bevii\ production.
 However,  \citet{McCollum2021ATel14655....1M}  found that  the progenitor object of V6595 Sgr is a cool subgiant with  T$_{eff}$ = 3750 $\pm$ 150 K, and log g = 3.5  $\pm$ 0.3, which is  unlikely   enriched either in   $^3$He or  $^4$He.

Evidence of  absorption components  have  always been observed in novae, both in hydrogen  and metal lines.
%such as  \nai\, \caii\ and \feii\. 
These absorptions  show  relatively   narrow
%(FWHM $\sim$ 0.6 $\textrm{\AA}$)  
 features named {\it THEA}, from \citet{Williams2008}. Generally, these are due to neutral and  low-ionization transitions of  elements that are  detected during the  optically thick phases of  classical novae. They exhibit only a single, low velocity absorption component, suggesting that they are located in a toroidal ejecta  surrounding the nova progenitor. These lines are  visible only for a limited time interval, generally few weeks after the nova discovery,  before disappearing and leaving a spectrum dominated by Balmer lines, Fe II, elements synthesised in the TNR 
 %like CNO elements 
 and, at later times  when the ionization state increases, by helium.
\citet{Aydi2020ApJ...905...62A} from a  study of their evolution in V906 Car argued that  they might be located in the slow ejecta component. 
The  detection of high-energy emission in gamma-rays  combined with the evidence of   delayed components in the optical  has challenged %posed new questions on   
the  nova phenomenology. \citet{Aydi2020ApJ...905...62A} suggested that  nova outburst  consists in   early dense toroidal ejecta moving outward with relatively slow ($ < 1000$ \kms) velocities which  produce  the first bright optical peak in the emission. Later,   a faster wind-like component   with larger velocities   will catch up with the earlier slower ejecta thus producing shocks that will be  observed at high-energies.
They concluded 
 that {\it THEA}  could be pre-existing material surrounding the system.  
 
 The \bevii\  absorption line observed in V6595 Sgr is expanding at almost 3000 \kms, which is the highest recorded velocity for \bevii\ in all novae. However, in V2944 Oph the velocity of the \bevii\  absorption is -645 \kms. {\b V5668 Sgr also shows \beviiii\ lines in a component at -730 \kms and also in components with velocities up to 2200 \kms.} By taking these figures at  face values,  \bevii\ is found in both high or low velocity components of nova outbursts.  This   shows  that TNR products are present also in the slower material.

\citet{Deenissenkov2021MNRAS.501L..33D}  noted that   the conditions after 71 minutes at the end of the  nova trajectory  implies a   small   occupation probabilities for K-shell electrons, suggesting that  \bevii\  is fully ionized and  practically stable.
They  computed the effective lifetime to be 
of the order of $5\times 10^6$ days.
%We note that if this is the case  it could also provide an alternative explanation for  the non detection of \liviii\ in  the outburst nova spectra.  \liviii\  was not seen to appear in  V5668 Sgr in  a time span of 100 days, about a factor two of  the \bevii\ laboratory decay time \citep{Molaro2016}. 
In Fig. \ref{fig:abundances}  the \bevii\ abundances both with and without the decay correction  are shown. When \bevii\  decay is not
taken into account the average value  is $ \log (N($\bevii$)/ N(HI)) +12 $  = $ 7.10 \pm 0.46 $.  The two sets of measurements show a  similar scatter  which exceeds the observational errors  and is  likely intrinsic.  The values without correction are significantly smaller  and     closer to the maximum theoretical values.
With this   scenario the distance between observed abundances and theoretical models is much reduced. 
%To   obtain   closer values through decay corrections   requires some fine tuning which is unlikely. 
V5668 Sgr is the only nova where two independent  measures of the \bevii\ abundance have been taken in two different epochs,  at 58 and 82 days after maximum,  respectively. The abundances differ by a factor of  1.8  when the decay is not taken into account, but  by only 30\%, i.e. within the measurement errors,  when the  decay is considered. Thus,  in this case  it seems that correction for the decay provides a more plausible behaviour.  Moreover, we note that the mere presence of the \beviiii\ lines in the outburst nova spectra is not consistent with \bevii\ remaining fully ionized for such a long time.

\section{Summary}

High resolution observations of the classical  nova V6595 Sgr taken with the UVES spectrograph   after few days from  explosion have been analysed to search for the presence of the \beviiii\ doublet lines. The main results are the following:

\begin{itemize}
    \item {The spectra showed the presence of a shallow absorption feature due to the \beviiii\ doublet lines spanning very high velocities from -2600 to -2900 \kms.  Superposed there are narrow absorption  components with velocities at $\sim$  -2620 and -2820 \kms. The absorptions are seen  in  several other elements such as \mgi, \nai\, \caii\, \feii\ and {H\,{\sc i}} as well.  The \bevii\ detection in   very high velocity ejecta components  shows that  both high and low velocity components contain material processed in the Termo-Nuclear Runaway. }
    \item{   Using \caii~ K line as a reference element, we infer    $X(\mbox{\bevii})/X(\mbox{H})$   $\approx$ 7.4 $\cdot 10 ^{-6}$, or  $\approx$ 9.8 $\cdot 10 ^{-6}$  when the \bevii\ decay is taken into account. The \bevii\ abundance  is about half of the value most frequently measured in novae. }
   
    \item{Observations 91 days after discovery showed  prominent emission lines of Neon and Oxygen which allowed to classify the nova as a  ONe type. \bevii\ seems to be produced in both types at comparable levels.}
    
     \item{ The possible presence of over-ionization suggested by \citet{chugai2020} in nova V5668 Sgr  is  discussed.   The model adopted by \citet{chugai2020} does indeed produce overionization in the observed layer, but, it fails to reproduce other observable and  does not reproduce the physical conditions of the material where the \bevii\ is observed. We therefore  conclude that  significant over-ionization  effects are    unlikely to occur. On the other hand, over-ionization could be present  for the fast  ONe novae resulting into a  \bevii\ abundance lower than CO novae, in better agreement with the  theoretical expectations.} 
    \item{The suggestion by \citet{Deenissenkov2021MNRAS.501L..33D}  that \bevii\   remains fully ionized and stable for a considerable time after the explosion is also discussed and  found  not consistent with the \beviiii\ observations and their time variation.}
%On the base of observations of V5668it has been suggested that the actual  \bevii\ abundance  depending on the presence of over-ionization in the absorbing components. Some considerations show    this  unlikely at the level  required    to reconcile observations with   current theoretical models.  

\end{itemize}

\begin{table}
\caption{   
A(\bevii)  
abundances    for the  novae with narrow
absorption components. 
The literature values are computed from the original  W(\beviiii,doublet)/W(CaII$_K$) by taking  the same 1D recommended solar A(Ca) = 6.34 $\pm$0.06  from \citep{lodders2019arXiv191200844L} but V838 Her where magnesium was used. 
$N($\bevii$)/N(H)_c$ are the values   corrected for the
 \bevii\ decay with a mean life of 76.8 days. References are: (1) \citet{Tajitsu2015}. Note that they did not correct for the \bevii\ decay; (2) \citet{Molaro2016},  (3) \citet{Izzo2018}, (4) \citet{Tajitsu2016}, (5) \citet{Selvelli2018}, the measurement from \mgii\ absorption is reported, (6) \citet{molaro2020MNRAS.492.4975M}, V612 Sct  could be  a peculiar object \citep{mason2020A&A...635A.115M}, (7) \citet{Arai2021ApJ...916...44A}, (8) \citet{Izzo2015},   the \livii\ abundance is measured  from   \liviii\ 670.8 nm  and ionization is estimated with the  \nai\, and   \ki\ lines, the latter is labeled with an asterisk, (9) this paper. }
\label{tab:3}
\begin{center}
\begin{tabular}{llrrrrrr}
\hline
\hline
\multicolumn{1}{c}{Nova} &
\multicolumn{1}{c}{type} &
\multicolumn{1}{c}{d} &
\multicolumn{1}{c}{comp} &
\multicolumn{1}{c}{ A(\bevii) } &
\multicolumn{1}{c}{ A(\bevii)$_c$ } &
\multicolumn{1}{c}{Ref}& \\
\multicolumn{1}{c}{} &
\multicolumn{1}{c}{} &
\multicolumn{1}{c}{} &
\multicolumn{1}{c}{\kms} &
\multicolumn{1}{c}{  } &
\multicolumn{1}{c}{} &
\multicolumn{1}{c}{}\\
\hline
V339 Del & CO & 47 & -1103& 6.92 &7.20  & 1,4 \\
V339 Del & CO &47  &-1268 &  7.11   & 7.38   & 1,4  \\
V5668 Sgr & CO &  58 & -1175 &  7.84 & 8.17   & 2   \\
 V5668 Sgr& CO & 82 &-1500 &7.58 &8.04  & 2  \\
 V2944 Oph  & CO &  80 &  -645 & 6.72  & 7.18  & 4 \\
V407 Lup  & ONe&  8  &-2030 & 7.69     & 7.73  & 3  \\
V838 Her & ONe?&  3  & -2500 &   7.66  & 7.68 &5   \\
%V838 Her & &  29  &  &  - &  20  & 5 \\
V612 Sct & ?&    &  & &  - & 6 \\
V357 Mus &CO? &  35  & $\approx$  -1000 & 6.96& 7.18&  6 \\
FM Cir & CO?&    & &   &: & 6 \\
V906 Car & CO?&  80  & $\approx$ -600 & 6.86 & 7.30& 6  \\
V5669 Sgr & CO & 28& $\approx$ -1000 &  6.34& 6.51 & 7  \\
V5669 Sgr & CO & 28& $\approx$ -2000 & 6.61 & 6.77& 7 \\
V6595 Sgr & ONe & 15  &-2700  & 6.87  & 6.99 &9 & \\
\hline
V1369 Cen & CO & 7& -550 & 5.00 & 5.04 & 8\\
V1369 Cen & CO & 13& -560 & 5.30 & 5.38 & 8\\
V1369 Cen$^*$ & CO & 7& -550  &4.70 & 4.78 & 8\\
V1369 Cen$^*$ & CO & 13& -560  &4.78 & 4.85 & 8\\
\hline
\end{tabular}
\end{center}

\end{table}

\section*{Acknowledgments}

 The ESO staff is warmly acknowledged for the execution of these observation  during the pandemic  lockdown. An anonymous referee is warmly thanked for many useful suggestions. Based on observations made with the Italian Telescopio Nazionale Galileo (TNG) operated on the island of La Palma by the Fundación Galileo Galilei of the INAF (Istituto Nazionale di Astrofisica) at the Spanish Observatorio del Roque de los Muchachos of the Instituto de Astrofisica de Canarias. LI was supported by two grants from VILLUM FONDEN (project number 16599 and 25501). LI warmly thanks Ernesto Guido for the technical support and important discussions during the planning of the observations.

\section{data availability}
 Based on  data from the UVES spectrograph at the Unit 2 of the VLT at the Paranal Observatory, ESO, Chile. The observations have been taken under a Target opportunity Program 105.20B6.001 PI P. Molaro.  ESO data are world-wide available and can be requested after the proprietary period of one year by the astronomical community through the link http://archive.eso.org/cms/eso-data.html. Before  they will be shared on  reasonable request to the corresponding author.

\bibliographystyle{mnras}
\bibliography{novae.bib}
  
\end{document}